\documentclass[a4paper,fleqn,usenatbib]{mnras}
\usepackage{mathptmx}
\usepackage[T1]{fontenc}
\usepackage{ae,aecompl}
\usepackage{times}
\usepackage{amssymb}
\usepackage{amsmath}
\usepackage{upgreek}
\usepackage{color}
\usepackage{graphicx}
\usepackage{pdflscape}

\title[JCMT Galactic Plane PGCCs]{A study of Galactic Plane \emph{Planck} Galactic Cold Clumps observed by SCOPE and the JCMT Plane Survey}
\author[D. J. Eden et al.]{D. J. Eden,$^{1}$\thanks{E-mail: david.eden@armagh.ac.uk} Tie Liu,$^{2}$ T.J.T. Moore,$^{3}$ J. Di Francesco,$^{4,5}$ G. Fuller,$^{6}$ Kee-Tae Kim,$^{7,8}$ \newauthor Di Li,$^{9,10}$ S.-Y. Liu,$^{11}$ R. Plume,$^{12}$ Ken'ichi Tatematsu,$^{13,14}$ M.A. Thompson,$^{15}$ Y. Wu,$^{16}$ \newauthor L. Bronfman,$^{17}$ H.M. Butner,$^{18}$ M.J. Currie,$^{19,20}$ G. Garay,$^{17}$ P.F. Goldsmith,$^{21}$ \newauthor N. Hirano,$^{11}$ D. Johnstone,$^{4,5}$ M. Juvela,$^{22}$ S.-P. Lai,$^{11,23,24,25}$ C.W. Lee,$^{7,8}$ \newauthor E.E. Mannfors,$^{22}$ F. Olguin,$^{25}$ K. Pattle,$^{26}$ Geumsook Park,$^{7}$ D. Polychroni,$^{27}$ \newauthor M. Rawlings,$^{28}$ A.J. Rigby,$^{15}$ P. Sanhueza,$^{29,30}$ A. Traficante,$^{31}$ J.S. Urquhart,$^{32}$ \newauthor B. Weferling,$^{33}$ G.J. White,$^{34,19}$ and R.K. Yadav,$^{35}$\\
Affiliations are listed at the end of the paper}

\date{Accepted XXX. Received YYY; in original form ZZZ}

\pubyear{2023}

\begin{document}
\label{firstpage}
\pagerange{\pageref{firstpage}--\pageref{lastpage}}
\maketitle

\begin{abstract}

We have investigated the physical properties of $\emph{Planck}$ Galactic Cold Clumps (PGCCs) located in the Galactic Plane, using the JCMT Plane Survey (JPS) and the SCUBA-2 Continuum Observations of Pre-protostellar Evolution (SCOPE) survey. By utilising a suite of molecular-line surveys, velocities and distances were assigned to the compact sources within the PGCCs, placing them in a Galactic context. The properties of these compact sources show no large-scale variations with Galactic environment. Investigating the star-forming content of the sample, we find that the luminosity-to-mass ratio ($L/M$) is an order of magnitude lower than in other Galactic studies, indicating that these objects are hosting lower levels of star formation. Finally, by comparing ATLASGAL sources that are associated or are not associated with PGCCs, we find that those associated with PGCCs are typically colder, denser, and have a lower $L/M$ ratio, hinting that PGCCs are a distinct population of Galactic Plane sources.

\end{abstract}


\section{Introduction}

Star formation occurs across the Galaxy, but a large fraction occurs within the Galactic Plane. Milky Way-wide surveys have revealed that the material needed to form stars is almost ubiquitous across the Galactic Plane, such as the molecular gas (e.g., \citealt{Dame01}) or the dust tracing the denser structures (e.g., \citealt{Molinari16}), along with the \emph{Planck} Galactic Cold Clumps (PGCCs; \citealt{Planck11,Planck16}). The $\emph{Planck}$ survey sought to map the cosmic microwave background, but in the process of doing so, also mapped the foreground emission. Whilst removing this emission, over 13\,000 PGCCs were identified across all Galactic environments, with a significant fraction ($\sim$\,20 per cent) within $\pm 2^{\circ}$ of the mid-plane and a further $\sim$\,20 per cent within $\pm 5^{\circ}$. Follow-up studies of these PGCCs have shown that, although they house the right physical conditions for star formation such as low dust temperatures \citep{Planck16}, CO \citep[e.g.][]{Zhang16}, dense-gas tracers such as HCN, HCO$^{+}$, N$_2$H$^+$ and NH$_{3}$ \citep[e.g.][]{Yuan16,Kim2020,Yi2021,Feher22,Berdikhan2024}, they have low levels of star-formation activity \citep{Tang18,Yi18,Zhang18} and are more quiescent and less evolved than the typical star-forming region \citep{Wu12,Liu13,Xu2024}. Adding to this, any correlation with young stellar objects tends to the youngest protostellar stages \citep{Juvela18}.

Recent follow-up observations with interferometers (e.g., ALMA) toward PGCCs in the Orion Giant Molecular Clouds (GMCs) have found that a large fraction of PGCCs contain centrally concentrated, high-density prestellar cores \citep{Hirano2024,Dutta2020,Sahu2021,Sahu2023} and/or young stellar objects with very collimated outflows \citep{Dutta2020,Dutta2022,Dutta2024,Jhan2022}. The cold and dense cores inside PGCCs also show high deuterium fractions of molecules \citep{Kim2020,Tatematsu2021}. These observational results indicate that PGCCs may represent the very early stages in star formation, albeit in a nearby star-forming region.

The physical state of the PGCCs thus makes them intriguing as a potential tracer of the earliest stages of star formation. This leads to two questions: are PGCCs significantly different from other star-forming Galactic Plane sources not associated with PGCCs, and do their properties vary as a function of Galactic environment? Galactic environments vary widely. Within the Galactic Centre, the conditions are much more extreme and akin to those in the early universe (i.e., $z \sim 2-3$; \citealt{Kruijssen13}), whilst in the Outer Galaxy there is a significantly lower metallicity \citep[e.g.,][]{Netopil22} and radiation field \citep[e.g.,][]{Popescu17}. Galactic-scale studies have indicated that, once a molecular cloud or dense clump forms, the star formation proceeds with the same average efficiency regardless of Galactic environment \citep[e.g.][]{Eden12,Eden13,Ragan18,Urquhart20,Eden21,Urquhart22}.  It is, also, important to know if these environmental differences can be detected in the physical properties of some of the youngest dust concentrations in our Galaxy, and whether these differences are imprinted in the ongoing star formation. 

This paper is organised as follows: Section 2 introduces the two James Clerk Maxwell Telescope (JCMT) surveys whose catalogues are used for the forthcoming analysis. Sections 3 and 4 introduce the radial-velocity and distance determinations, respectively, whilst Section 5 contains the Galactic distribution of the PGCCs. Sections 6 and 7 address the physical properties of the whole sample and the star-forming content, respectively. Section 8 addresses the differences in the PGCC sample compared to other Galactic Plane sources by utilising the ATLASGAL survey \citep{Schuller09}, and Section 9 summaries the work and states our conclusions.

\section{SCUBA-2 Data}

The Galactic Plane PGCCs observed at the JCMT with Submillimetre Common-User Bolometer Array 2 (SCUBA-2; \citealt{Holland13}) in the 850-$\upmu$m continuum are contained within two surveys. The first 174 PGCCs were observed as part of the JCMT Plane Survey (JPS; \citealt{Moore15,Eden17}), which is one of the JCMT Legacy Surveys \citep{Chrysostomou10}. The JPS was a targeted, yet unbiased, survey of the Galactic Plane, observing equally spaced regions covering approximately 5$\degr$\,$\times$\,1.7$\degr$ centred at $\ell$\,=\,10$\degr$, 20$\degr$, 30$\degr$, 40$\degr$, 50$\degr$, and 60$\degr$. The SCUBA-2 Continuum Observations of Pre-protostellar Evolution (SCOPE; \citealt{Liu18,Eden19}) survey observed PGCCs in the Galactic Plane outside of the JPS regions. The SCOPE survey observed 204 Galactic Plane PGCCs.

Another JCMT Legacy Survey, the SCUBA-2 Ambitious Sky Survey (SASSy; \citealt{MacKenzie11,Nettke17}; Thompson et al, in preparation) has observed the Galactic Plane with SCUBA-2. However, its rms sensitivity does not reach the required threshold of 6\,mJy\,beam$^{-1}$, which was the desired rms of the SCOPE survey, and, as such, SASSy is not used in this work.

\begin{figure*}
\begin{center}
\includegraphics[width=\textwidth]{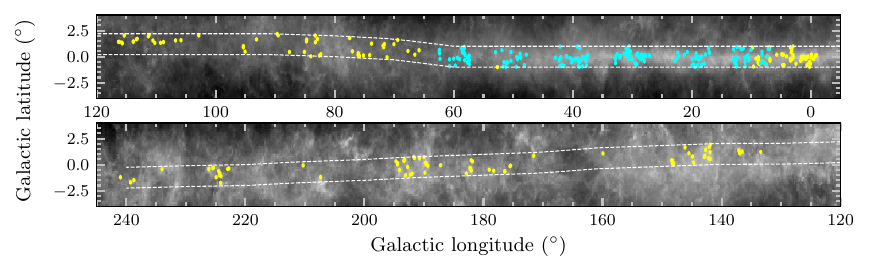}
\caption{The observed PGCCs in the Galactic Plane. The cyan circles represent the sources from the JPS \citep{Eden17}, whilst the yellow circles are those observed within the SCOPE survey \citep{Eden19}. The white dashed lines are the extent of the Galactic Plane considered, as derived from the Hi-GAL survey \citep{Schisano20}. The background image is the $\emph{Planck}$ dust opacity map \citep{Planck14}.}
\label{pgccs}
\end{center}
\end{figure*}

In this study, we have only considered PGCCs that were found within the latitude range of the $\emph{Herschel}$ infrared Galactic Plane Survey (Hi-GAL; \citealt{Molinari10,Molinari16}). Hi-GAL followed the warp of the Galaxy \citep{Schisano20}. The distribution of the observed PGCCs is displayed in Fig.~\ref{pgccs}.

\subsection{Compact source extraction}

Compact sources were extracted from within each of the JPS and SCOPE surveys using the {\sc FellWalker} (FW; \citealt{Berry15}) algorithm. A full explanation of the processes used in each survey can be found in \citet{Eden17,Eden19}. JPS and SCOPE contained 7813 and 3528 compact sources, respectively.

Due to the targeted nature of the SCOPE survey, the PGCCs were observed using the CV Daisy mode of SCUBA-2 \citep{Bintley14}, compared to the pong3600 maps used in the JPS. The CV Daisy mode is most suitable for compact sources, and produced an rms of 6 mJy\,beam$^{-1}$ in the central-most 3 arcmin, and an rms comparable to the pong3600 maps in the JPS out to a radius of 6 arcmin (43.9 mJy\,beam$^{-1}$ in SCOPE compared to 25--31 mJy\,beam$^{-1}$ in JPS; \citealt{Eden17,Eden19}). Therefore, we extracted all compact sources in the JPS catalogue within 6 arcmin of any catalogued PGCC \citep{Planck11,Planck16}. This accounted for 1447 compact sources associated with 148 PGCCs, with 26 PGCCs undetected. The Galactic Plane PGCCs from the SCOPE catalogue contained 1731 compact sources, distributed across 169 PGCCs with a 35 undetected PGCCs.

\section{Radial-velocity determination}
\label{sec:radial}

To determine the distance to each PGCC, and therefore Galactic environment and physical properties, the radial velocity of the source compared with the local standard of rest ($v_{\rmn{lsr}}$) is required. This velocity can then be compared with a Galactic rotation model \citep[e.g.][]{Brand93,Reid14}. The velocities for the Galactic Plane PGCCs are available in a suite of molecular-line Galactic Plane surveys, namely the Galactic Ring Survey (GRS, $^{13}$CO $J=1-0$; \citealt{Jackson06}); the CO Heterodyne Inner Milky
Way Plane Survey (CHIMPS, $^{13}$CO/C$^{18}$O $J=3-2$; \citealt{Rigby16}); CHIMPS2 ($^{12}$CO/$^{13}$CO/C$^{18}$O $J=3-2$; \citealt{Eden20}); the CO High Resolution Survey (COHRS, $^{12}$CO $J=3-2$; \citealt{Dempsey13,Park23}); the FOREST Unbiased Galactic Plane Imaging Survey (FUGIN, $^{12}$CO/$^{13}$CO/C$^{18}$O $J=1-0$; \citealt{Umemoto17}); Structure, excitation, and dynamics of the inner Galactic interstellar medium (SEDIGISM, $^{13}$CO/C$^{18}$O $J=2-1$; \citealt{Schuller17}); and Milky Way Imaging Scroll Painting (MWISP, $^{12}$CO/$^{13}$CO/C$^{18}$O $J=1-0$; \citealt{Su19}).

\begin{figure*}
\begin{center}
\includegraphics[width=\textwidth]{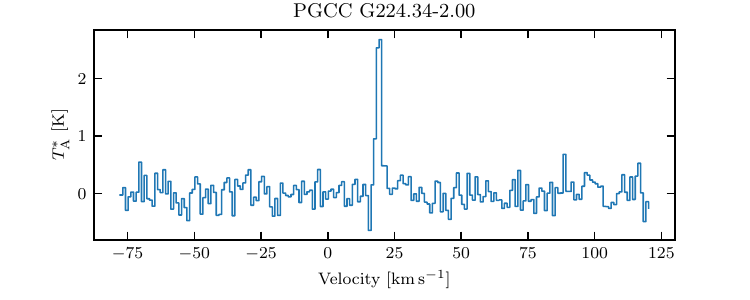}
\caption{Example spectrum for velocity determination. This spectrum is for PGCC G224.34-2.00 and is the $^{13}$CO $J=3-2$ line from the CHIMPS2 survey (Eden et al., in preparation). The velocity determined for this source was 19.3 km\,s$^{-1}$.}
\label{spectrum}
\end{center}
\end{figure*}

To obtain these velocities, the spectra were inspected at the position of the relevant PGCC. All the compact sources extracted from the JPS and SCOPE surveys \citep{Eden17,Eden19} are assumed to have the same velocity as the associated PGCC. This process was completed for all surveys that had coverage at the position of a PGCC. The spectra were inspected in decreasing critical-density order. The isotopologue order was C$^{18}$O, $^{13}$CO, $^{12}$CO, with a further breakdown via rotational transition, with $J=3-2$, then $J=2-1$, and finally $J=1-0$. If multiple emission peaks are present in the spectrum, the strongest emission peak was chosen under the assumption that this would correspond to the highest column density along that line of sight \citep[e.g.,][]{Urquhart07,Urquhart09,Eden12,Eden13}. An example spectrum is displayed in Fig.~\ref{spectrum}, displaying the $^{13}$CO $J=3-2$ line for the PGCC G224.34-2.00 from the CHIMPS2 survey (\citealt{Eden20}; Eden et al., in preparation).

Using these surveys resulted in 273 of the 317 PGCCs having assigned velocities. The surveys used for each are listed in Table~\ref{velocities}. The remaining 44 were assigned velocities using different methods. The first was to positionally cross-match the remaining sources with the ATLASGAL-survey catalogue, which contains velocities from a wide range of sources in the literature and a self-contained survey (full details can be found in \citealp{Urquhart18}). This resulted in a further 16 allocated velocities. The remaining 28 velocities were assigned by extracting the spectra from the composite $^{12}$CO  $J=1-0$ survey of \citet*{Dame01}. The data sets used from the composite survey are the first- and second-quadrant surveys of \citet{Dame01}, the survey of Cygnus X \citep{Leung92}, and the third-quadrant survey of \citet{May93}.

\begin{table*}
\centering
\caption{Surveys used to assign radial velocities to the 317 PGCCs.}
\label{velocities}
\begin{tabular}{lcccll}
\hline
Survey & Total & JPS & SCOPE & Molecular & References \\
 & PGCCs & PGCCs & PGCCs & Transition & \\
\hline
CHIMPS & 17 & 7 & 10 & $^{13}$CO/C$^{18}$O ($J=3-2$) & [1] \citealt{Rigby16} \\
CHIMPS2 & 3 & 0 & 3 & $^{12}$CO/$^{13}$CO/C$^{18}$O ($J=3-2$) & [2] \citealt{Eden20} \\
COHRS & 67 & 34 & 33 & $^{12}$CO ($J=3-2$) & [3] \citealt{Park23} \\
SEDIGISM & 8 & 6 & 2 & $^{13}$CO/C$^{18}$O ($J=2-1$) & [4] \citealt{Schuller21} \\
GRS & 27 & 20 & 7 & $^{13}$CO ($J=1-0$) & [5] \citealt{Jackson06} \\
FUGIN & 89 & 52 & 37 & $^{12}$CO/$^{13}$CO/C$^{18}$O ($J=1-0$) & [6] \citealt{Umemoto17} \\
MWISP & 62 & 0 & 62 & $^{12}$CO/$^{13}$CO/C$^{18}$O ($J=1-0$) & [7] \citealt{Su19} \\
ATLASGAL & 16 & 13 & 3 & $^{13}$CO/C$^{18}$O ($J=2-1$) & [8] \citealt{Urquhart18} \\
DHT Survey & 23 & 16 & 7 & $^{12}$CO ($J=1-0$) & [9] \citealt{Dame01} \\
Cygnus X & 4 & 0 & 4 & $^{12}$CO ($J=1-0$) & [10] \citealt{Leung92} \\
Third Quadrant & 1 & 0 & 1 & $^{12}$CO ($J=1-0$) & [11] \citealt{May93} \\
\hline
\end{tabular}
\end{table*}

\begin{figure*}
\begin{center}
\includegraphics[width=\textwidth]{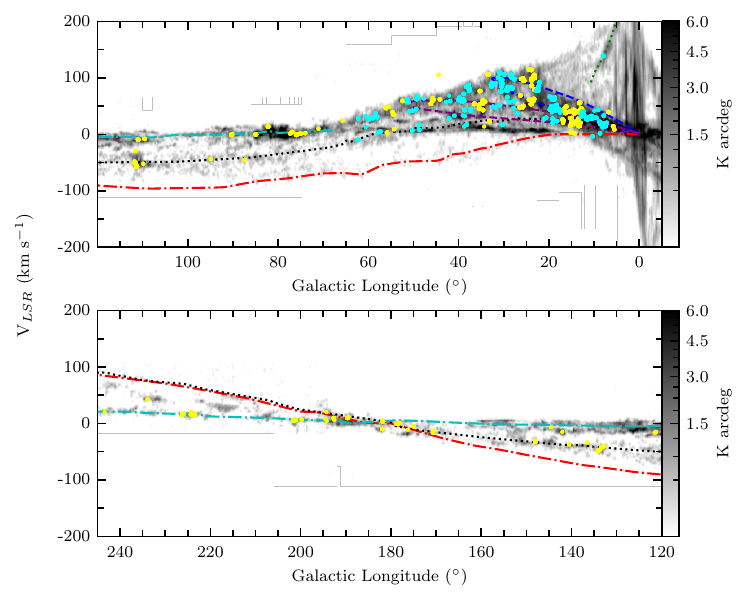}
\caption{Longitude-velocity distribution of the PGCCs with detected JPS (cyan; \citealt{Eden17}) and SCOPE (yellow; \citealt{Eden19}) compact sources. The sources are plotted on top of the $^{12}$CO $J=1\rightarrow0$ emission from the \citet{Dame01} survey. The loci of the spiral arms from \citet{Reid19} are also plotted. The Scutum--Centaurus Arm is the blue dashed line, the Sagittarius Arm is represented by the purple dashed line, the Perseus arm is the black dotted line, the red dot-dash line is the Outer Arm and the cyan dashed line is the Local Arm. The Connecting Arm is also displayed by the green dotted line.}
\label{l-v_pgccs}
\end{center}
\end{figure*}

The longitude-velocity $(\ell-V_{\rmn{LSR}})$ diagram of the PGCCs with a detected JPS \citep{Eden17} and/or SCOPE \citep{Eden19} source is displayed in Fig.~\ref{l-v_pgccs}. The sources are overlaid on the $^{12}$CO $J=1\rightarrow0$ integrated emission from \citet{Dame01}. This emission traces the global Galactic distribution of molecular gas, and the PGCCs are well correlated with this emission. Also overlaid on the $^{12}$CO $J=1\rightarrow0$ emission map are the loci of the spiral arms from \citet{Reid19}. The PGCCs are also tightly correlated with the spiral arms. The Galactic distribution of the PGCCs, and their relation to Galactic structure will be addressed in Section~\ref{sec:Galactic}.

\section{Distance Determination}

Determining the heliocentric distance to the PGCCs, and the compact sources within their substructure, is vital to calculating many physical properties of the sources. We have employed a multi-stage process to assign distances to all 317 observed PGCCs and, therefore, the 3178 associated compact sources. The steps in this process are outlined in Section~\ref{sec:stagesdistance}, with the summary of distances assigned displayed in Table~\ref{distancemethods} and are in the order of reliability.

The velocities described in Section~\ref{sec:radial} are employed to determine kinematic distances to the majority of sources within this sample. However, sources within the Solar Circle (Galactocentric radius of $<$ 8.15\,kpc \citealt{Reid19}) are subject to kinematic distance ambiguity (KDA), giving two potential distances to each source. Steps (iii)--(ix) in Section~\ref{sec:stagesdistance} relate to resolving this KDA, and assigning a unique distance to each source.

The rotation curve used to give Galactocentric radius ($R_{\rmn{GC}}$) measurements here is that of \citet{Brand93}. The choice of rotation curve is not vital, as available rotation curves agree within the errors in the distances \citep[e.g.,][]{Eden12}, which are assumed to be of the order of 30 per cent \citep{DuarteCabral21}.

\begin{table}
\centering
\caption{Summary of methods used to assign distances, with the number of PGCCs and compact sources determined at each step. The roman numerals in the first column relate to the steps detailed in Section~\ref{sec:stagesdistance}.}
\label{distancemethods}
\begin{tabular}{llcc}
\hline
Step & Method & No. of & Associated \\
& & PGCCs assigned & compact sources \\
\hline
(i) & Maser parallax & 6 & 99\\
(ii) & Outer Galaxy & 79 & 422\\
(iii) & Tangent velocity & 18 & 141\\
(iv) & Scale height & 54 & 493\\
(v) & ATLASGAL match & 79 & 1187\\
(vi) & \ion{H}{i} SA Near & 34 & 379\\
(vii) & \ion{H}{i} SA Far & 7 & 47\\
(viii) & IRDC & 25 & 291\\
(ix) & Bayesian method & 15 & 119\\
\hline
\end{tabular}
\end{table}

\subsection{Methodology}
\label{sec:stagesdistance}

\subsubsection{(i) Maser parallax}

We matched the position of the PGCCs to the maser parallax catalogue of \citet{Reid19}, with a positional match (8 arcmin) in longitude, latitude, and velocity (km\,s$^{-1}$) allowing this distance to be adopted. This method resulted in distances to 6 PGCCs, with a total of 99 compact sources.

\subsubsection{(ii) Outer Galaxy sources}

Sources located at Galactocentric radii greater than 8.15\,kpc have no KDA, and therefore the kinematic distance can be assigned. 79 PGCCs were assigned distances using this method, associated with 422 compact sources.

\subsubsection{(iii) Tangent velocity}

The KDA gives two equally spaced distance solutions about a tangent point. Those sources whose $V_{\rmn{LSR}}$ is within 10\,km\,s$^{-1}$ of the tangent velocity are assigned the tangent distance. The two KDA solutions in this situation are closer together than any uncertainties associated with the distance-determination process (30 per cent; \citealt{DuarteCabral21}). A total of 18 PGCCs, with 141 compact sources, were assigned distances using this method.

\subsubsection{(iv) Scale-height distribution}

Studies have shown that high-mass star-forming regions are found towards the Galactic mid-plane, with a scale height of 30\,pc \citep[e.g.][]{Reed00}. We have set a tolerance for the scale height of the far-distance solution of four times this, 120\,pc, following previous studies \citep[e.g.][]{Urquhart11,Urquhart18}. If the far kinematic distance solution of a PGCC results in a projected scale height that is larger than this tolerance, it is placed at the near distance solution. This accounts for 54 and 493 PGCCs and compact sources, respectively.

\subsubsection{(v) ATLASGAL comparisons}

The remaining 160 sources were compared to the catalogue of the ATLASGAL survey, and their associated distances and velocities \citep{Urquhart18}. These sources were matched positionally (8 arcmin tolerance), and any PGCCs that had a matching velocity within 10\,km\,s$^{-1}$ had the ATLASGAL distance assigned to it. This method allowed for 79 PGCCs to be given distances, with an associated 1187 compact sources.

\subsubsection{(vi) \ion{H}{i} self-absorption, near distance}

The HISA, \ion{H}{i} self-absorption, method \citep[e.g.][]{Roman-Duval09} compares the spectrum of \ion{H}{i} emission to the radial velocity measured for a given source. If at the near distance, an absorption feature would be present in the \ion{H}{i} spectrum as the cold \ion{H}{i} embedded in a molecular cloud would absorb the emission from the warmer, background \ion{H}{i} that is ubiquitous in the interstellar medium. This absorption feature would be coincident with the velocity measured for the PGCC. We compared the \ion{H}{i} data from the VLA Galactic Plane Survey \citep{Stil06} and the Galactic All-Sky Survey \citep{McClure-GAriffiths09,Kalberla10}, and found absorption features coincident with the measured velocities of 34 PGCCs, placing them at the near distance. These 34 PGCCs have 379 compact sources associated with them.

\subsubsection{(vii) HISA far distance}

Of the 47 PGCCs that did not have an absorption feature present in the \ion{H}{i}, the remaining spectra had two potential solutions. The first is that the PGCC velocity was coincident with a peak in the \ion{H}{i} spectrum. This is because the embedded \ion{H}{i} is behind the background emission, and therefore, there is no absorption. The second was that it was an ambiguous feature, with a full explanation of this sort of feature found in \citet{Urquhart18} and \citet{DuarteCabral21}. 7 PGCCs were assigned to the far distance, with 47 compact sources given those distances.

\subsubsection{(viii) IRDC comparisons}

The existence of an infrared dark cloud (IRDC) against the background infrared emission implies that a source is in the foreground, or at the near distance. A study of IRDCs found that 89 per cent were found to be at the near distance \citep{Giannetti15}, therefore, if the PGCCs were coincident with an IRDC, we have placed them at the near distance. We positionally matched the remaining PGCCs without assigned distances to the IRDC catalogue of \citet{Peretto09}, within 8 arcmins, and found 25 that could be placed at the near distance. This also accounted for 291 compact sources.

\subsubsection{(ix) Bayesian distance}

A total of 15 PGCCs, and 119 compact sources, were left without distance assignment. We used the Bayesian distance model of \citet{Reid16,Reid19} which uses the positions of the spiral arms, derived from the maser parallaxes. This model preferentially places sources in the spiral arms, which is why it is not used to derive the distance for all PGCCs in this study. However, when all other available methods have been exhausted, we use this method.

\subsection{Distance summary}

We have derived distances to the 317 PGCCs, and therefore 3178 compact sources, from the SCOPE and JPS surveys within the Galactic Plane. These distances were assigned using the nine steps listed above. A tenth method was also used between steps (v) and (vi), comparing \ion{H}{i} absorption features towards \ion{H}{ii} regions and comparisons to the \ion{H}{ii} region catalogues of \citet{Kolpak03}, \citet{Anderson09} and \citet{Urquhart12} resulted in no assigned distances. The \ion{H}{ii} region catalogues of \citet{Urquhart13} and \citet{Wienen15} were not used as they were based on ATLASGAL data, and these distances would be accounted for in step 5. A small portion of the velocities, distances, and methods used for the individual PGCCs are displayed in Table~\ref{tab:PGCCvelocities}. The full table is available from the Supporting Information. From this point forward, the compact sources JPS and SCOPE will be treated as one sample.

\begin{table}
\centering
\begin{minipage}{\linewidth}
\caption{Distance and velocity information for the PGCCs. The number in the $V_{\rmn{LSR}}$ reference column relates to the numbers in the reference column of Table\,\ref{velocities}, whilst the Distance Method column relates to the steps in Table\,\ref{distancemethods}.}
\label{tab:PGCCvelocities}
\begin{tabular}{lccccc}
\hline
PGCC/Region & $V_{\rmn{LSR}}$ & $V_{\rmn{LSR}}$ & $R_{\rmn{GC}}$ & Distance & Distance \\
 & (km\,s$^{-1}$) & Reference & (kpc) & (kpc) & Method \\
\hline
G28.38+0.10	&	105.4	&	3	&	4.2	&	5.7	&	(viii)	\\
G28.48+0.21	&	100.0	&	3	&	4.3	&	5.8	&	(v)	\\
G28.56$-$0.24	&	82.5	&	3	&	4.7	&	4.7	&	(v)	\\
G28.94$-$0.04	&	56.0	&	3	&	5.5	&	3.4	&	(vi)	\\
G29.18+0.24	&	76.0	&	3	&	4.9	&	4.3	&	(viii)	\\
G29.25$-$0.71	&	82.4	&	6	&	4.7	&	4.5	&	(v)	\\
G29.28$-$0.77	&	64.8	&	6	&	5.2	&	3.8	&	(iv)	\\
G29.31+0.17	&	81.0	&	3	&	4.8	&	4.5	&	(vi)	\\
G29.60$-$0.62	&	77.1	&	6	&	4.9	&	4.4	&	(v)	\\
G30.02$-$0.27	&	104.9	&	1	&	4.3	&	7.6	&	(v)	\\
G30.34+0.48	&	15.9	&	6	&	7.3	&	13.2	&	(v)	\\
G30.48+0.56	&	90.1	&	6	&	4.6	&	5.0	&	(viii)	\\
G30.52$-$0.11	&	89.2	&	1	&	4.6	&	4.9	&	(vi)	\\
G30.52+0.99	&	95.3	&	6	&	4.5	&	5.3	&	(iv)	\\
G30.55+0.16	&	83.6	&	6	&	4.8	&	5.0	&	(ix)	\\
\hline
\end{tabular}
$\emph{Notes:}$ Only a small portion of the data are displayed here, with the full table available from the Supporting Information.
\end{minipage}
\end{table}

\begin{figure*}
\begin{center}
\includegraphics[width=\textwidth]{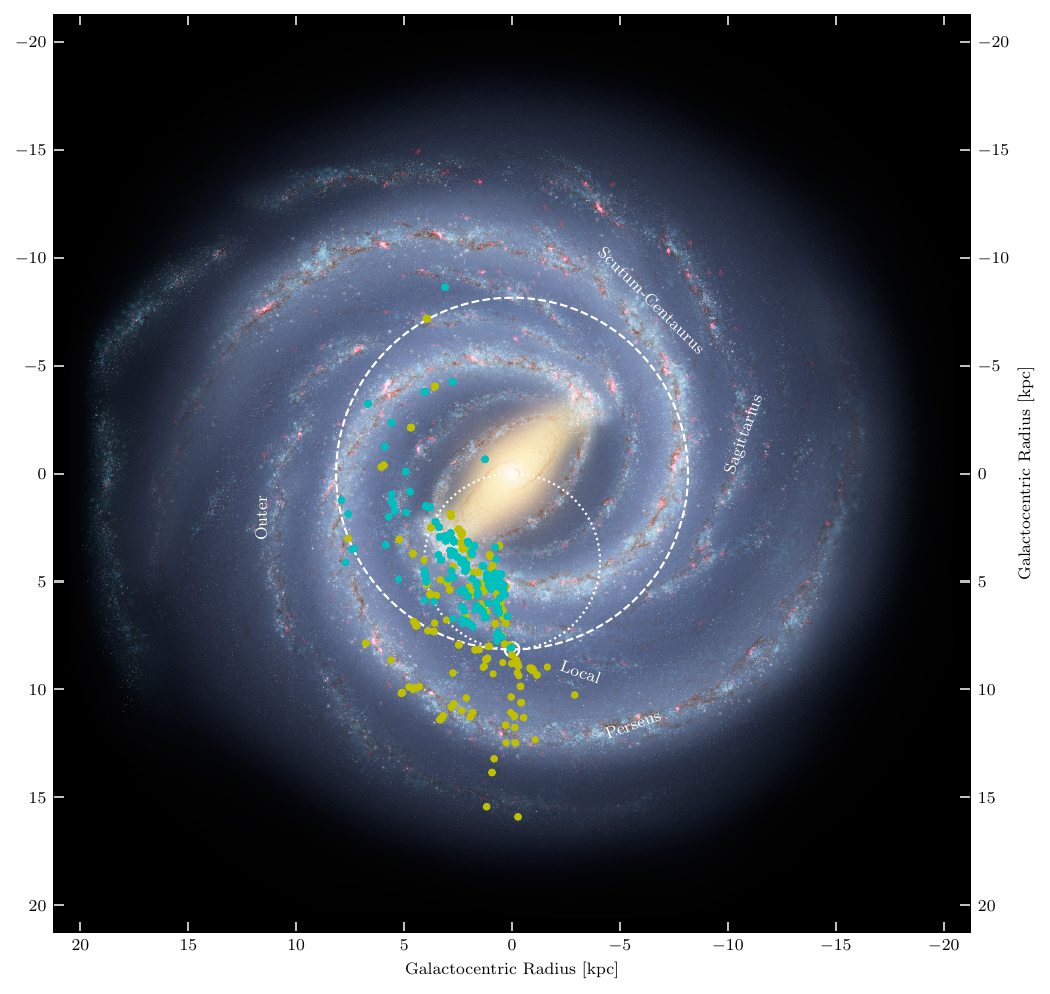}
\caption{Face-on view of the distribution of the 317 PGCCs (JPS in cyan, SCOPE in yellow) using the kinematic distances derived here. They are overlaid on the schematic of the Milky Way produced by Robert Hurt of the Spitzer Science Center, in consultation with Robert Benjamin (University of Wisconsin-Whitewater). The position of the Sun is indicated by the white solar symbol, with the white dotted line representing the tangent positions and the white dashed line showing the extent of the Solar Circle. We have also indicated the positions of the four major spiral arms, and the Local Arm.}
\label{faceon}
\end{center}
\end{figure*}

\begin{figure}
\begin{center}
\includegraphics[width=0.5\textwidth]{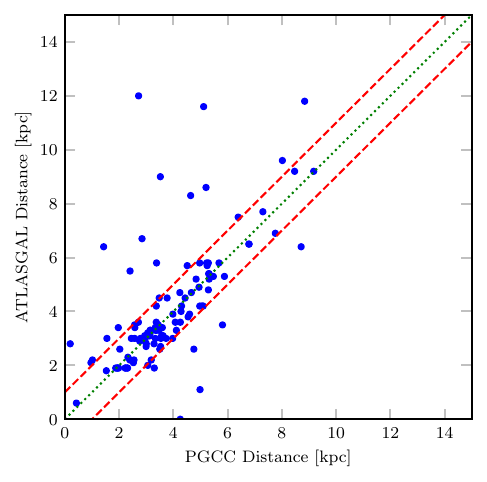}
\caption{Distance comparison of PGCC sources in the JPS and SCOPE surveys with the compact sources of the ATLASGAL survey \citep{Urquhart18}. The green dotted line represents the 1:1 line, whereas the red dashed lines are the 1\,kpc limits.}
\label{atlascomparison}
\end{center}
\end{figure}

The distribution of the 317 PGCCs with derived distances is shown in Fig.\,\ref{faceon}. We have also positionally matched the PGCC sources not derived using method $\emph{(v)}$ above with compact sources from the ATLASGAL survey \citep{Urquhart18} in order to compare distance determinations. This comparison is shown in Fig.~\ref{atlascomparison}. There are 115 matches within an 8-arcmin search radius, with 88 (77 per cent) of these having a distance that can be considered to be consistent, i.e., within 1\,kpc. 22 of the remaining 27 sources have the ATLASGAL source placed at a larger distance to that of the PGCC. However, although impacting the derived quantities of radius, mass, and mass surface density of individual sources, incorrect distance assignments are unlikely to affect ensemble properties \citep{Rani23}.

\section{Galactic Distribution}
\label{sec:Galactic}

In Figs.\,\ref{l-v_pgccs} and \,\ref{faceon} we show the results of the velocity and distance analysis presented in the previous sections. The positions of the sources are well correlated with the $^{12}$CO emission, as well as the spiral arms from the \citet{Reid19} model. The correlation with the spiral arms is also clear from the face-on image of the Galaxy in Fig.\,\ref{faceon}.

Fig.\,\ref{l-v_pgccs} is split into two segments, the first covers $\ell = -5\degr$ to $120\degr$ and includes the four major spiral arms, the Scutum--Centaurus, Sagittarius, Perseus, and Outer arms. It also includes the Local Arm and the Connecting Arm. The Connecting Arm is thought to be the nearside dust lane around the Central Molecular Zone (CMZ) that are signs of accretion onto the CMZ \citep{Sormani19}. The Connecting Arm appears to have one source associated with it, whilst the majority of the other sources can be found in the inner 30$\degr$ of the Galaxy and associated with the Scutum--Centaurus and Sagittarius arms. There are also a significant number of sources found in the $\ell-V_{\rmn{LSR}}$ space between these two arms, a region where a connecting spur is identified \citep{Stark06,Rigby16}. The Outer Arm does not appear to have any sources associated with it, away from the central 10$\degr$ of the Galaxy, although the distance to the Outer Arm in this line of sight is large, which may lead to an observational bias.

In the second segment, covering $\ell = 120\degr$ to $240\degr$, three spiral arms are found, the Perseus, Outer, and Local arms. Again in this segment, the spiral arms trace the positions of the PGCCs well. The $\ell-V_{\rmn{LSR}}$ spaces occupied by the Perseus and Outer arms are indistinguishable at longitudes greater than $\ell = 180\degr$. However, referring to the derived distances of the sources that occupy that $\ell-V_{\rmn{LSR}}$ space, two sources are found to be in the Outer Arm, as can be seen in Fig.~\ref{faceon}.

\begin{figure*}
\begin{tabular}{ll}
\includegraphics[width=0.49\textwidth]{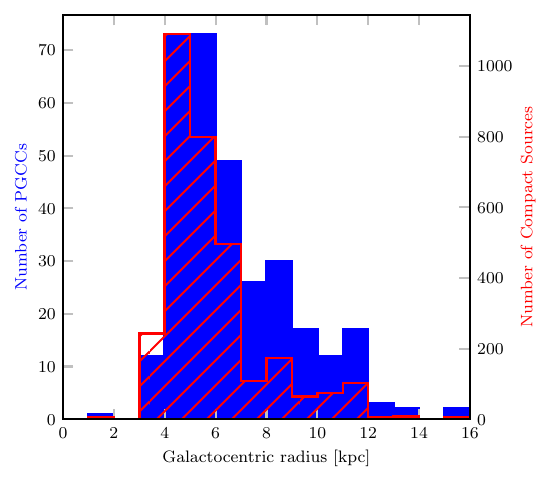} & \includegraphics[width=0.49\textwidth]{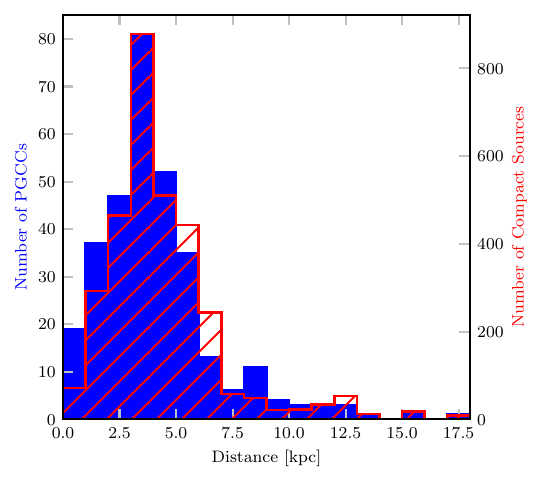} \\
\end{tabular}
\caption{Distribution of PGCCs (blue, solid histogram) and the associated compact sources (red, hashed histogram) as a function of Galactocentric radius and helicoentric distance in the left and right panels, respectively.}
\label{dist_distribution}
\end{figure*}

The distribution of the PGCCs, and the associated compact sources, as a function of Galactocentric and helicoentric distance are displayed in Fig.~\ref{dist_distribution}. These two distributions reflect what is displayed in Figs~\ref{l-v_pgccs} and \ref{faceon}, with a peak in the Galactocentric radii distribution found between 4 and 7\,kpc, corresponding to the collection of sources found in the Scutum--Centaurus and Sagittarius arms. Further, smaller peaks are found at 8 and 12\,kpc, which can be assigned to the first and second Galactic quadrant portions of the Perseus Arm, respectively. The heliocentric distances display a peak at approximately 3.5\,kpc, which includes sources both in the Inner Galaxy and in the Perseus arm in the Galactic anti-centre. There are also a significant number of sources found at distances of $\sim$\,5\,kpc, corresponding to the Scutum--Centaurus and Sagittarius arms in the Inner Galaxy.

By using the spiral arm loci in $\ell-V_{\rmn{LSR}}$ space from \citet{Reid19} and the derived Galactocentric distances, we determined the nearest spiral arms to each PGCC. If within 10\,km\,s$^{-1}$, the PGCC is assumed to be associated with that arm. However, any PGCCs found to be more than 10\,km\,s$^{-1}$ away from any spiral arm are considered to be interarm sources. This velocity cut is used as 95 per cent of SEDIGISM clouds are found to be within this tolerance of the nearest spiral arm \citep{Urquhart20}.

\section{Physical Properties}
\label{sec:physical}

We calculate a series of physical properties for the compact sources, using the distances derived above and the flux information from the catalogues in \citet{Eden17} and \citet{Eden19}.For those sources with positionally matched ATLASGAL sources, the ATLASGAL temperatures are used. All other sources follow the distribution of temperatures as a function of Galactocentric radius in the ATLASGAL survey \citep{Urquhart18}. A small portion of the calculated physical properties for the compact sources are contained in Table~\ref{physicalprops} with the full table available from the Supporting Information.

\begin{table*}
\begin{center}
\caption{Derived compact-source parameters.}
\label{physicalprops}
\begin{tabular}{llcccccccc}\hline
Source Name & Region & $\ell_{\rmn{peak}}$ & $b_{\rmn{peak}}$ & $V_{\rmn{LSR}}$ & Distance & Radius & log$[N_{\rmn{H_{2}}}]$ & log$[$Mass Surface & log$[M_{\rmn{clump}}]$ \\
& & & & & & & & Density$]$ & \\
& & ($^{\circ}$) & ($^{\circ}$) & (km\,s$^{-1}$) & (kpc) & (pc) & (cm$^{-2}$) & (M$_{\odot}$\,pc$^{-2}$) & (M$_{\odot}$)\\
\hline
SCOPEG005.83$-$00.94	&	G005.91$-$01.00	&	5.831	&	$-$0.935	&	13.0	&	2.95	&	0.37	&	22.901	&	3.018	&	2.657	\\
SCOPEG005.84$-$01.00	&	G005.91$-$01.00	&	5.838	&	$-$0.995	&	13.0	&	2.95	&	0.93	&	23.045	&	2.866	&	3.301	\\
SCOPEG005.85$-$00.99	&	G005.91$-$01.00	&	5.853	&	$-$0.993	&	13.0	&	2.95	&	0.50	&	22.893	&	2.960	&	2.857	\\
SCOPEG005.87$-$00.99	&	G005.91$-$01.00	&	5.873	&	$-$0.994	&	13.0	&	2.95	&	0.57	&	22.757	&	2.875	&	2.888	\\
SCOPEG005.88$-$00.94	&	G005.91$-$01.00	&	5.879	&	$-$0.943	&	13.0	&	2.95	&	0.83	&	22.683	&	2.790	&	3.127	\\
SCOPEG005.89$-$00.91	&	G005.91$-$01.00	&	5.885	&	$-$0.914	&	13.0	&	2.95	&	0.56	&	22.543	&	2.712	&	2.704	\\
SCOPEG005.89$-$00.94	&	G005.91$-$01.00	&	5.892	&	$-$0.944	&	13.0	&	2.95	&	0.46	&	22.923	&	2.879	&	2.698	\\
SCOPEG005.90$-$00.93	&	G005.91$-$01.00	&	5.904	&	$-$0.931	&	13.0	&	2.95	&	0.44	&	22.749	&	2.797	&	2.589	\\
SCOPEG005.91$-$00.95	&	G005.91$-$01.00	&	5.914	&	$-$0.951	&	13.0	&	2.95	&	0.73	&	23.143	&	2.889	&	3.113	\\
SCOPEG005.92$-$00.99	&	G005.91$-$01.00	&	5.917	&	$-$0.990	&	13.0	&	2.95	&	0.89	&	23.072	&	2.994	&	3.388	\\
SCOPEG005.92$-$00.96	&	G005.91$-$01.00	&	5.923	&	$-$0.955	&	13.0	&	2.95	&	0.32	&	22.543	&	2.714	&	2.207	\\
SCOPEG005.92$-$00.97	&	G005.91$-$01.00	&	5.924	&	-0.967	&	13.0	&	2.95	&	0.42	&	22.526	&	2.676	&	2.410	\\
SCOPEG006.91$+$00.88	&	G006.9$+$00.8A1	&	6.914	&	0.879	&	38.4	&	4.84	&	0.80	&	22.373	&	2.535	&	2.836	\\
SCOPEG006.94$+$00.91	&	G006.9$+$00.8A1	&	6.940	&	0.910	&	38.4	&	4.84	&	0.54	&	22.547	&	2.829	&	2.789	\\
SCOPEG006.94$+$00.92	&	G006.9$+$00.8A1	&	6.940	&	0.921	&	38.4	&	4.84	&	1.27	&	22.539	&	2.637	&	3.339	\\
\hline
\multicolumn{10}{l}{$\emph{Note:}$ Only a small portion of the catalogue is shown here. The entire catalogue is available in the Supporting Information.}\\
\end{tabular}
\end{center}
\end{table*}

\subsection{Masses}
\label{sec:masses}

The JCMT compact source masses are calculated using the optically thin conversion:

\begin{equation}
M_{\rmn{clump}} = \frac{S_{\nu}D^{2}}{\kappa_{\nu}B_{\nu}(T_{\rmn{d}})},
\end{equation}

\noindent where $S_{\nu}$ is the 850-$\upmu$m integrated flux density, $D$ is the distance to the source, $\kappa_{\nu}$ is the mass absorption coefficient taken to be 0.01\,cm$^{2}$\,g$^{-1}$, whilst also accounting for the gas-to-dust ratio of 100 \citep{Mitchell01}; and $B_{\nu}(T_{\rmn{d}})$ is the Planck function evaluated at $T_{\rmn{d}}$, where $T_{\rmn{d}}$ is the temperature derived from the matching ATLASGAL source or the fit of average clump temperatures as a function of Galactocentric radius from figure 12 in \citet{Urquhart18}, with a median temperature of 19.8\,K.

\begin{figure}
\begin{tabular}{l}
\includegraphics[width=0.49\textwidth]{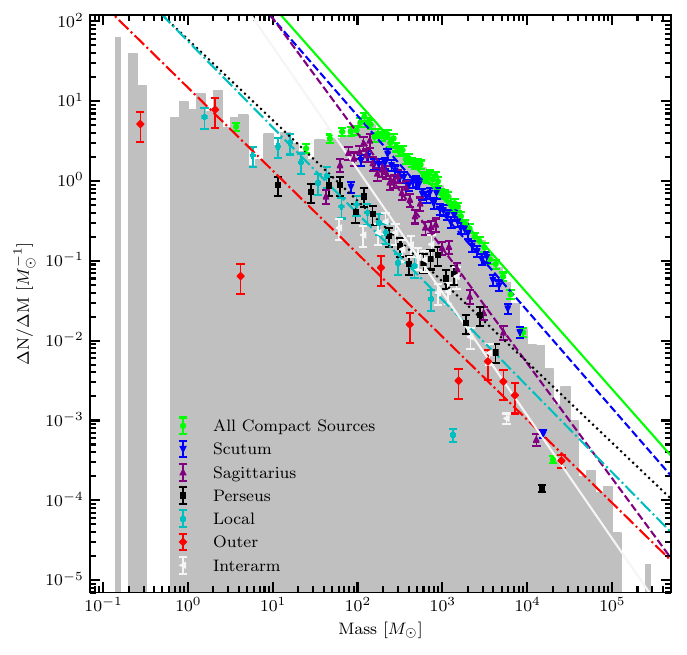}\\
\includegraphics[width=0.49\textwidth]{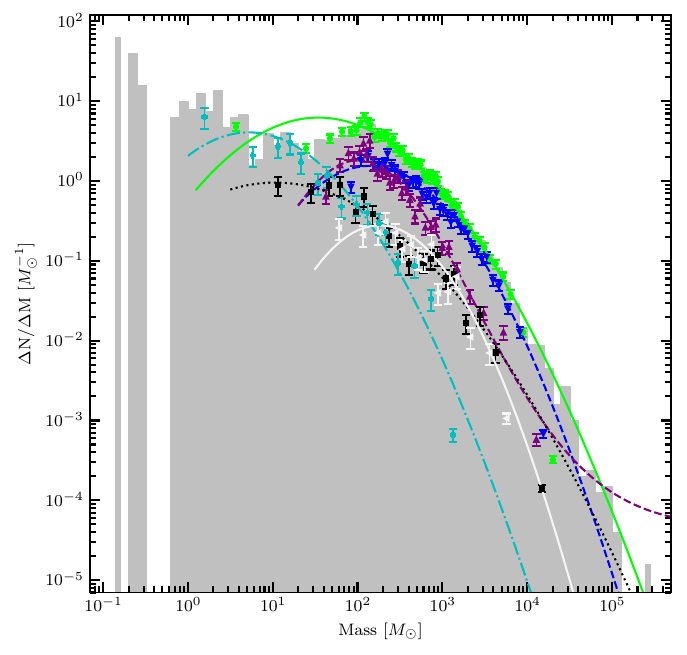}\\
\end{tabular}
\caption{The clump-mass distributions of all the compact sources in the Galactic Plane PGCCs, along with the mass distributions of the different spiral arms. The grey histogram is the entire compact source catalogue. The points represent the mass distributions as described in the text, using equally populated bins. The green circles are for the entire catalogue, the downwards-pointing blue triangles are the Scutum--Centaurus Arm, the upwards-pointing purple triangles are the Sagittarius Arm, the black squares are the Perseus Arm, the cyan circles are the Local Arm, with the red diamond and left white triangles representing the Outer Arm and Interarm sources, respectively. The fits for the total sample, Scutum--Centaurus, Sagittarius, Perseus, Local, Outer arms, and Interarm are indicated by the solid green, dashed blue, dashed purple, dotted black, dot-dash cyan, dot-dash red, and white solid lines, respectively. The top panel displays the least-squares fits to these distributions, assuming the form of a power law, whilst the bottom panel indicates log-normal fits to the distributions. The sample for the Outer Arm is omitted from the log-normal distributions since a fit was not possible to that sample.}
\label{massfunction}
\end{figure}

The distribution of masses of the compact sources is displayed in Fig.~\ref{massfunction}. The plotted quantity is the bin population per unit mass interval, $\Delta\emph{N/}\Delta\emph{M}$, with the histogram of the entire catalogue using equal bin widths shown in grey.  The superimposed points show the mass distributions using equal bin populations, to equalise the weights from Poisson errors \citep{MaizApellaniz05}.  These are separated into the entire catalogue, the sources associated with individual spiral arms and the interarm sources. The last point positions are calculated as in \citet{Eden15} and \citet{Eden18}, with the mass coordinate represented by the median value in each bin.

By assuming a power-law slope of the form $\Delta\emph{N/}\Delta\emph{M}$ $\propto$ $M^{\alpha}$, a least-squares fit can be applied above the completeness limits  to calculate the values of the indices, $\alpha$. The completeness limits were determined as the most-massive bin that follows two bins of lower $\Delta\emph{N/}\Delta\emph{M}$ values, or in the case of the Outer arm, which does not fit this criterion, the highest value of 
$\Delta\emph{N/}\Delta\emph{M}$. The least-squares fit uses an outlier-resistant method to ensure that the last bin, which is potentially quite wide, does not influence the fit unduly \citep{Eden18}. The values for the indices, and the corresponding completeness limits (the lower end of the fitting range) are given in Table~\ref{massindices}.

The index values for the individual spiral arms are all consistent within 3$\upsigma$ with the total sample, except for the Sagittarius arm and the Interarm regions, which are steeper by more than $\sim$\,5$\upsigma$. The Sagittarius Arm has been postulated as being different from the two ``major'' spiral arms: Perseus and Scutum--Centaurus \citep{Benjamin08}. However, it is unclear if this would be reflected in the clump-mass function, since there is no evidence of a difference in its star-formation properties \citep{Urquhart14a}.

The physics of the interarm regions may influence the steeper index found for these clumps. Modelling indicates molecular clouds within a spiral potential are more massive \citep{Dobbs11}. Observations and simulations have indicated that there is a statistical link between the stellar initial mass function and clump-mass distributions \citep[e.g.][]{Simpson08,Pelkonen21} with an altered initial mass function found (IMF) in more-massive clouds \citep{Weidner10}. This altered IMF tends to be more top-heavy, i.e., flatter, and is caused by the suppression of fragmentation due to radiative feedback \citep{Krumholz11}, which reflects the conditions found in spiral arms compared with interarm regions \citep{Koda12}. The excitation temperatures, linewidths, and virial parameters of interarm clouds are also found to be lower in the interarm regions \citep{Rigby19}. This is taken to be a result of the lack of external pressure that is applied from the ambient arm material in the spiral arms, which allows the clouds with higher virial parameters to disperse in the interarm. This breakup of molecular clouds would allow for less opportunity for clumps to form.

Along with power-law fits, mass functions have also been found to be well described by log-normal distributions \citep[e.g.][]{Peretto10,Wienen15}. We fit each sample, the total and the individual spiral arms, with an adapted version of the expression outlined in \citet{Peretto10} in which the mass function has the form $\Delta\emph{N/}\Delta\emph{M} = Ae^{-x}$ where:

\begin{equation}
x = \frac{(\rmn{log}_{10}M - \rmn{log}_{10}M_{\rmn{peak}})^{2}}{2\sigma^{2}}
\end{equation}

\noindent In this form, $A$ is a constant, $M$ is the bin mass, $M_{\rmn{peak}}$ is the mass at the peak of the distribution, and $\sigma$ is the dispersion of the distribution. The values for $M_{\rmn{peak}}$ and $\sigma$ are presented in Table~\ref{massindices}, with the fits displayed in the lower panel of Fig.~\ref{massfunction}. No value is presented for the Outer Arm as a log-normal fit did not converge for that sample. The completeness limits and values for $M_{\rmn{peak}}$ are in good agreement for the Scutum--Centaurus and Sagittarius arms, along with the interarm sample. The remaining distributions, the total sample, and Perseus and Local arms have $M_{\rmn{peak}}$ values below the inferred completeness limit, indicating the peaks are not well constrained. However, for all samples, other than the aforementioned Outer Arm distribution, they are better described as log-normal as opposed to power-law fits, as indicated by the $\chi^{2}$ values for each fit, as shown in Table~\ref{massindices}.

\begin{table*}
\centering
\caption{Fit parameters for the mass distributions of the entire compact-source sample, and those for the individual spiral arms. The indices of the power-law fits, the completeness limits, and therefore, the lower end of the fitting range for these indices are also shown. The log-normal peaks and dispersions are listed, along with the $\chi^{2}$ for each fit.}
\label{massindices}
\begin{tabular}{lcccccc}
\hline
Spiral Arm & Index & Completeness & $\chi^{2}$ & M$_{\rmn{peak}}$ & $\sigma$ & $\chi^{2}$ \\
& & Limit (M$_{\odot}$) & Power Law & (M$_{\odot}$) & (log$[$M$_{\odot}$$]$) & Log Normal \\
\hline
Total & $-1.20 \pm 0.03$ & 120 & 378.50 & 35 & 6.81 & 4.09 \\
Scutum--Centaurus & $-1.22 \pm 0.05$ & 220 & 249.76 & 140 & 4.88 & 1.48 \\
Sagittarius & $-1.44 \pm 0.04$ & 110 & 86.95 & 100 & 1.38 & 2.80 \\
Perseus & $-1.01 \pm 0.07$ & 60 & 56.84 & 11 & 8.36 & 1.79 \\
Local & $-1.08 \pm 0.05$ & 95 & 86.74 & 5.2 & 5.44 & 10.14 \\
Outer & $-1.04 \pm 0.06$ & 2.0 & 50.52 & & & \\
Interarm & $-1.54 \pm 0.15$ & 210 & 63.73 & 180 & 4.07 & 0.99 \\
\hline
\end{tabular}
\end{table*}

\subsection{Radii, and column and surface densities}

\begin{figure}
\begin{tabular}{l}
\includegraphics[width=0.3\textheight]{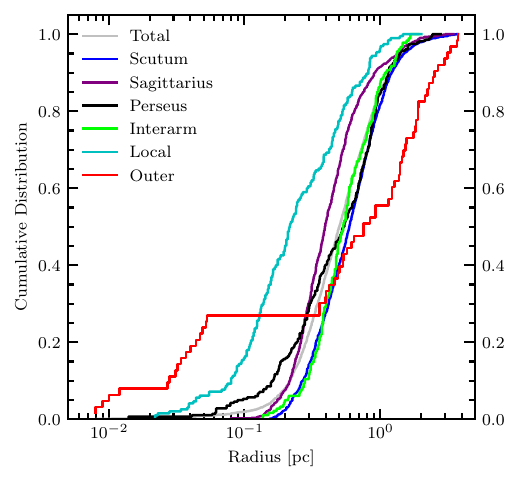} \\ \includegraphics[width=0.3\textheight]{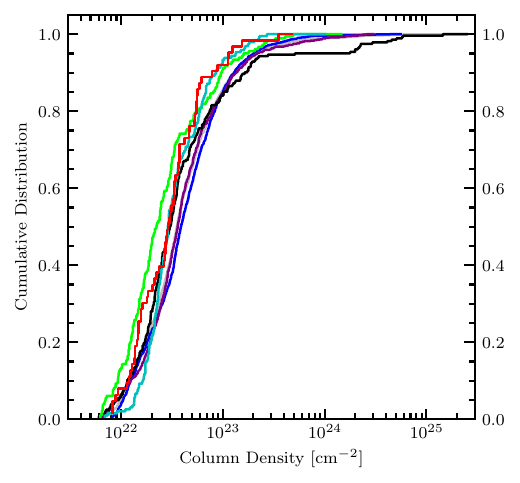} \\
\includegraphics[width=0.3\textheight]{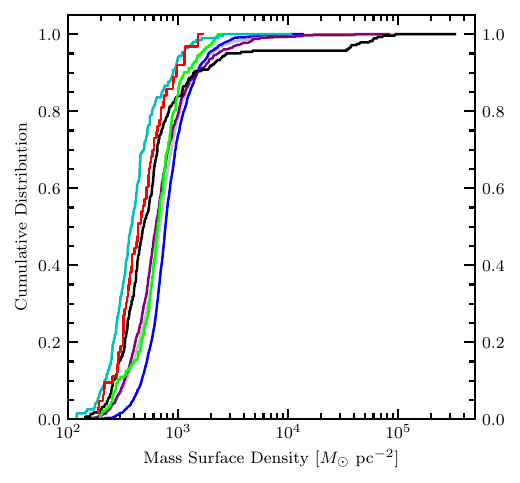} \\
\end{tabular}
\caption{Cumulative distributions of compact source radii, peak column densities, and surface densities in the top, middle, and bottom panels, respectively. The total sample, Scutum--Centaurus, Sagittarius, Perseus, Local, Outer, and Interarm samples are represented by the grey, blue, purple, black, cyan, red, and lime lines, respectively.}
\label{radii+densities}
\end{figure}

The cumulative distributions of the compact source radii, column densities and surface densities are shown in the top, middle, and bottom panels of Fig.~\ref{radii+densities}, respectively. These distributions are split into sources associated with the individual spiral arms.

The radii were calculated using the effective radii of the compact sources, as reported in \citet{Eden17} and \citet{Eden19} and the distances calculated above.

The column densities were calculated using the following equation:

\begin{equation}
N_{\rmn{H_{2}}} = \frac{S_{\nu, \rmn{peak}}}{B_{\nu}(T_{\rmn{d}})\Omega_{\rmn{b}}\kappa_{\nu}m_{\rmn{H}}\upmu}
\end{equation}

\noindent where $B_{\nu}(T_{d})$ and $\kappa_{\nu}$ are as defined in the previous section, $\Omega_{b}$ is the solid angle of the beam from a full-width half maximum of 14.4\,arcsec \citep{Eden17}, $m_{\rmn{H}}$ is the mass of a hydrogen atom, and $\upmu$ is the mean mass per hydrogen molecule, taken to be 2.8 \citep{Kauffmann08}. $S_{\nu, \rmn{peak}}$ is the peak flux density, where the SCOPE fluxes are converted from mJy\,arcsec$^{-2}$ to Jy\,beam$^{-1}$. The mass surface densities ($M_{\rmn{clump}}/\pi R^{2}$) make use of the masses and radii for each compact source.

The statistics of these distributions are shown in Table~\ref{parameterstats}, with the values for individual compact sources presented in Table~\ref{physicalprops}. The column density and mass surface densities are largely consistent with each other, which is not unsurprising since these quantities are fairly insensitive to heliocentric distance \citep{Dunham11}. However, differences in the mass surface density of molecular clouds are found between spiral arms and interarm environments \citep{Colombo22}. Since these differences are not reflected in the compact sources formed within these molecular clouds, it is more evidence to support conclusions drawn from Galactic Plane studies, that once a compact source forms within a molecular cloud, the environment in which it resides has very little impact on its properties \citep[e.g.,][]{Eden13,Urquhart15,Urquhart18,Urquhart22}, since no environmental variations are found. The column densities have statistics similar to those found from previous studies of SCUBA-2 PGCCs, with consistent mean values \citep[e.g.,][]{Mannfors21}.

The mean radii of compact sources in each spiral arm are not statistically consistent with each other, with Anderson--Darling (A--D) tests finding that they are largely not drawn from the same sample. The Local and Outer arms are the biggest outliers (A--D $p$-values $\ll$ 0.001), with the radii of Local Arm sources significantly lower, and the Outer Arm sources larger. This is not surprising due to the heliocentric distances involved, and previous studies finding the classification of sources detected varying with distance \citep[e.g.][]{Dunham11}.

\begin{table}
\centering
\caption{Statistics for physical parameters in the compact-source sample and the individual spiral-arm subsets. The mean ($\bar{x}$), standard error, standard deviation ($\sigma$), and median values are given. The first row in each section describes the entire sample.}
\label{parameterstats}
\begin{tabular}{lcccc}
\hline
Parameter/Spiral Arm & $\bar{x}$ & $\frac{\sigma}{\sqrt{N}}$ & $\sigma$ & Median \\
\hline
{\bf Radius (pc)} & 0.63 & 0.01 & 0.47 & 0.50\\
Scutum--Centaurus & 0.70 & 0.01 & 0.46 & 0.60 \\
Sagittarius & 0.51 & 0.01 & 0.40 & 0.39\\
Perseus & 0.63 & 0.03 & 0.46 & 0.55\\
Local & 0.34 & 0.02 & 0.31 & 0.22\\
Outer & 1.15 & 0.13 & 1.07 & 0.85\\
Interarm & 0.64 & 0.03 & 0.35 & 0.54\\
\hline
{\bf log$[N_{\rmn{H_{2}}}$ (cm$^{-2}$)$]$} & 22.60 & 0.01 & 0.43 & 22.55\\
Scutum--Centaurus & 22.62 & 0.01 & 0.40 & 22.09\\
Sagittarius & 22.61 & 0.01 & 0.43 & 22.56\\
Perseus & 22.61 & 0.04 & 0.59 & 22.49\\
Local & 22.53 & 0.02 & 0.31 & 22.46\\
Outer & 22.49 & 0.05 & 0.36 & 22.48\\
Interarm & 22.44 & 0.03 & 0.42 & 22.35\\
\hline
{\bf log$[$Mass Surface Density (M$_{\odot}$\,pc$^{-2}$)$]$} & 2.91 & 0.00 & 0.28 & 2.84\\
Scutum--Centaurus & 2.91 & 0.01 & 0.21 & 2.89 \\
Sagittarius & 2.83 & 0.01 & 0.30 & 2.80\\
Perseus & 2.79 & 0.03 & 0.44 & 2.69\\
Local & 2.61 & 0.02 & 0.26 & 2.58 \\
Outer & 2.65 & 0.03 & 0.22 & 2.63\\
Interarm & 2.82 & 0.02 & 0.22 & 2.82\\
\hline
\end{tabular}
\end{table}

\section{Star formation in PGCCs}

\subsection{Prestellar and protostellar compact sources}

Previous observations have found that the star formation in PGCCs is low compared with other star-forming regions in the Milky Way \citep[e.g.,][]{Tang18,Yi18,Zhang18}. To determine if a compact source from the sample in this study is hosting a young stellar object (YSO), a positional match within 40-arcsec was made with the band-merged catalogue of the Hi-GAL survey \citep{Elia17,Elia21}, where the presence of a Hi-GAL band-merged source containing a 70-$\upmu$m source indicated that it was protostellar following the methodology used in a number of studies \citep[e.g.,][]{Ragan16,Elia17,Ragan18,Elia21}. This resulted in a sample of 2107 protostellar compact sources that will be used to calculate star-formation efficiencies in Section~\ref{sec:sfe} using their Hi-GAL-determined luminosity \citep{Elia17,Elia21}. The luminosities were scaled using the Hi-GAL-derived distances \citep{Mege21} and the distances in this study. As a result of this matching, a total of 1071 prestellar compact sources were identified.

\subsection{Properties of proto- and prestellar compact sources}

We can now compare the physical properties derived in Section~\ref{sec:physical} for the protostellar and prestellar compact source subsamples.

\begin{figure}
\begin{tabular}{l}
\includegraphics[width=0.49\textwidth]{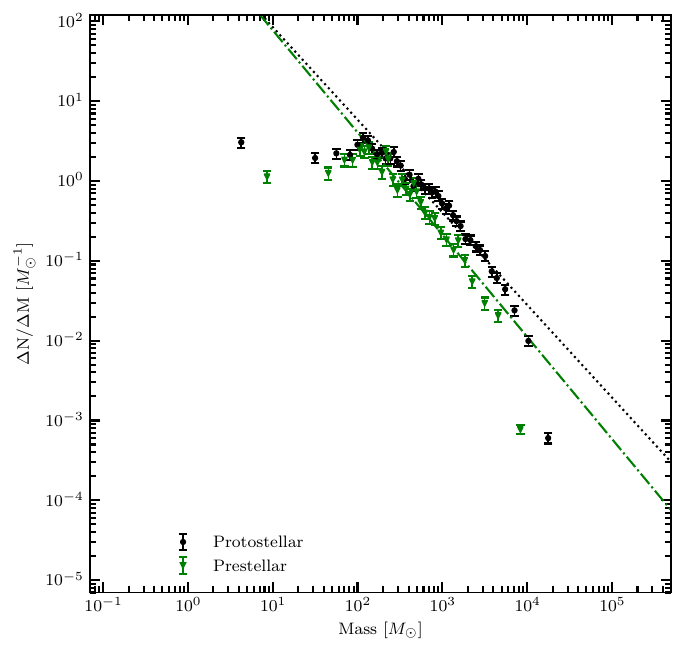}\\
\includegraphics[width=0.49\textwidth]{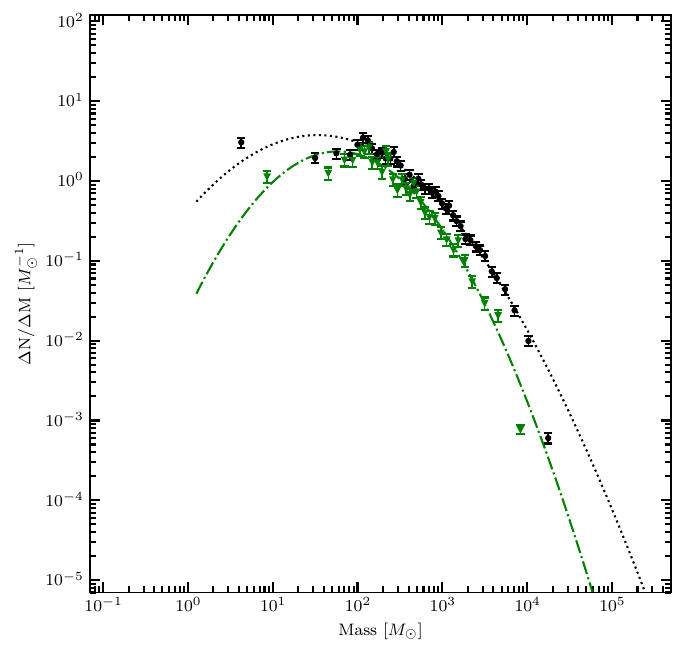}\\
\end{tabular}
\caption{The clump-mass distributions of the protostellar (black circles) and prestellar (green triangles) compact sources. The fits are represented by the black dotted and green dot-dashed lines, respectively with the least-square fits in the top panel, on the assumption of a power law, and the bottom panel displaying log-normal fits.}
\label{starforming_mass}
\end{figure}

The mass functions are displayed in Fig.~\ref{starforming_mass}. The indices of the protostellar and prestellar mass functions are found to be $\upalpha = -1.16 \pm 0.04$ and $\upalpha = -1.28 \pm 0.05$, respectively. These indices and peak values are consistent with each other, a result not matching that of the Hi-GAL survey \citep{Elia17}. The mass ranges of the two samples are also consistent, something also not observed in Hi-GAL \citep{Elia17}. It is, however, consistent with a previous PGCC result in \citet{Mannfors21}. The improved angular resolution of the JCMT compared  with \emph{Herschel} may be the cause of this, something speculated on by \citet{Elia17}. The mass functions presented here are also not split by distance, and are averaged over all Galactic environments, whereas the Hi-GAL study split their mass functions into 0.5\,kpc heliocentric distance bins. By combining multiple distances here, we may have averaged out these differences. As with the mass distributions of the individual spiral arms, we also fit log-normal distributions to the mass functions. The $\chi^{2}$ values for the power-law fits were found to be 28.87 and 25.01 for the protostellar and prestellar mass functions, respectively, whilst for the log-normal fits were 2.94 and 2.97, respectively. As with the earlier mass distributions, these samples are best described by a log-normal distribution.

\begin{figure}
\begin{tabular}{l}
\includegraphics[width=0.3\textheight]{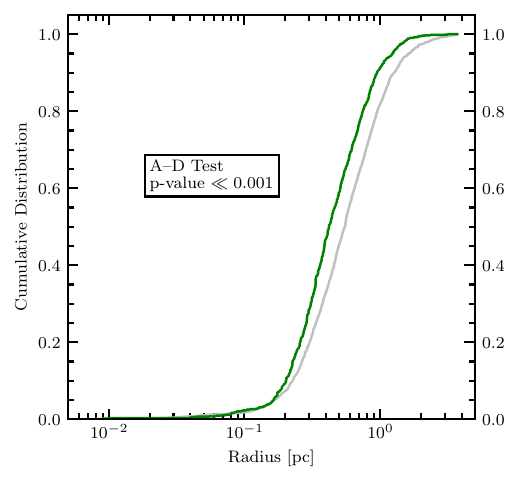} \\ \includegraphics[width=0.3\textheight]{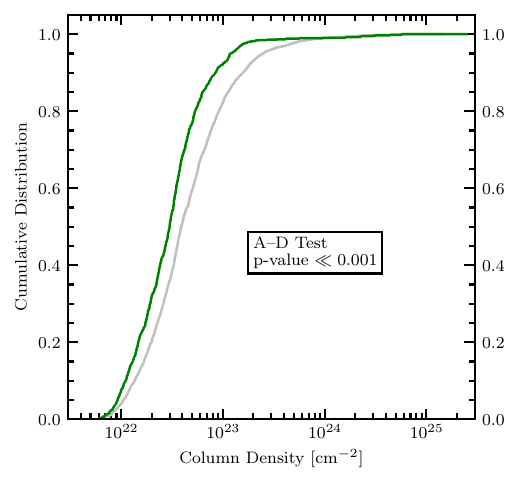} \\
\includegraphics[width=0.3\textheight]{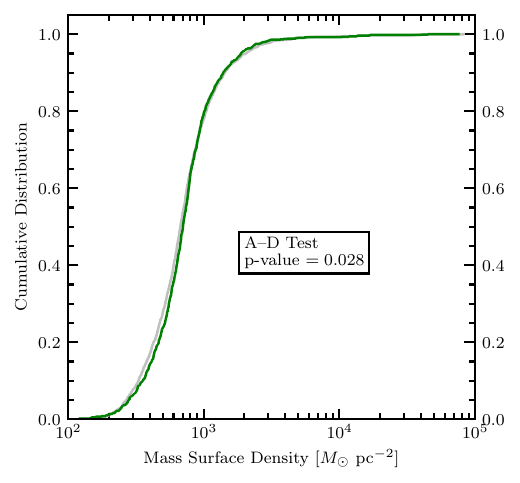} \\
\end{tabular}
\caption{Cumulative distributions of the compact source radii (top), column densities (middle), and surface densities (bottom) for the protostellar (grey) and prestellar (green) samples.}
\label{starforming_physical}
\end{figure}

The cumulative distributions of the radii, column densities, and surface densities for the protostellar and prestellar subsamples are displayed in Fig.~\ref{starforming_physical}. A--D tests were performed on each set and these found that the null hypothesis that the radii and column densities of protostellar and prestellar clumps are drawn from the same population can be rejected at a probability of $\ll 0.001$. Whereas this cannot be rejected in the mass surface densities, which found a A--D $p$-value of 0.028, or 2.2-$\upsigma$. The ATLASGAL survey split their protostellar sample into 70-$\upmu$m-bright, mid-IR bright, and massive star forming. Differences were found between prestellar and star-forming sources as traced by massive-star formation indicators \citep{Urquhart18}, although when samples complete in mass and column density were used, their differences largely disappeared \citet{Billington19,Urquhart22}.

\subsection{Mass-radius relationship}

\begin{figure*}
\includegraphics[width=\textwidth]{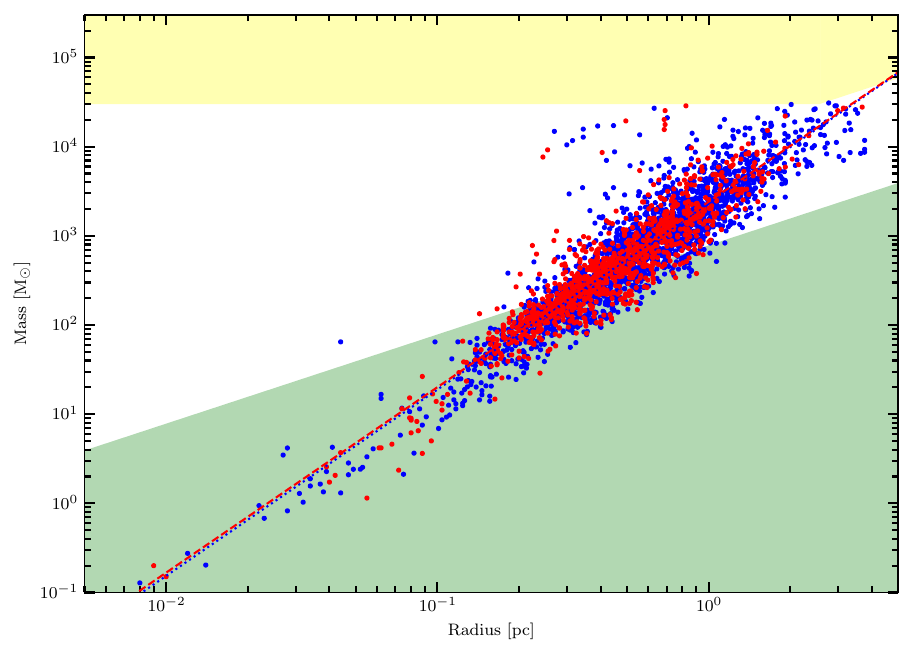}
\caption{The mass-radius relationship for the SCOPE compact sources. The protostellar sources are indicated by the blue circles, and the prestellar represented by the red circles. The blue dotted and red dashed lines represent least-squares linear fits to the proto and prestellar samples with slopes of $2.089\pm0.019$ and $2.083\pm0.028$, respectively. The green region represents the parameter space where massive star formation does not occur, satisfying the relationship $m(r) \leq 580 \rmn{M_{\odot}}(R_{\rmn{eff}}/\rmn{pc})^{1.33}$ \citep{Kauffmann10}. The yellow region at the upper end of the plot is where young massive clusters would be found \citet{Bressert12}.}
\label{mass-radius}
\end{figure*}

The relationship between the mass and size of a molecular cloud was initially described by \citet{Larson81} indicating that a constant column density, with respect to radius, was found. Further work found that different slopes were required to describe sources that are star forming, and within which mass regime that star formation was occurring \citep{Kauffmann10a}.

The mass-radius relationship for the SCOPE compact sources is displayed in Fig.~\ref{mass-radius}. The sample is again split into protostellar and prestellar sources. Both subsamples show that these two parameters are well correlated, with Pearson correlation coefficients of 0.92 and 0.91, respectively. Linear least-squares fitting gives slopes of $2.089 \pm 0.019$ and $2.083 \pm 0.028$ for the protostellar and prestellar subsamples, respectively. This is consistent with the relationship found by the initial $\emph{Planck}$ PGCC analysis \citep{Planck16}, but it is steeper than the relationship found in the ATLASGAL survey of $1.647 \pm 0.012$ \citep{Urquhart18}. The steepness of this relationship may point towards the more-quiescent nature of PGCCs, even in those that are forming stars. The slopes found here are consistent with the relationship found in quiescent molecular clouds in a tidal dwarf galaxy, whilst the ATLASGAL slope is consistent with the star-forming molecular clouds in that particular galaxy \citep{Querejeta21}.

Fig.~\ref{mass-radius} is separated into three regions, each with a different significance. The green shaded area is where low-mass star formation would occur, whilst high-mass star formation would be expected to occur in the unshaded regions. This delineation is defined by the relationship $m(r) = 580\, \rmn{M_{\odot}}(R_{\rmn{eff}}/\rmn{pc})^{1.33}$, below which would be mainly low-mass star formation \citep{Kauffmann10}. The final region, indicated in yellow, is where young massive clusters would form \citep{Bressert12}. The majority (88.5 per cent) of PGCC compact sources are found in the high-mass star-forming space, with this fraction identical in both prestellar and protostellar sources. Notably, no young massive cluster-forming sources ($M \gtrsim 10^{4}$\,M$_{\odot}$, younger than 100\,Myr; \citealt{PortegiesZwart10}) are found in this study, either protostellar or prestellar. It is not surprising that no starless examples were found, since numerous studies have also failed to find any in the Galactic disc \citep[e.g.][]{Ginsburg12,Longmore17,Urquhart18}.

There are a number of compact sources on the edge of the cluster-forming source region, namely SCOPEG015.01$-$00.67, SCOPEG035.58$+$00.05, SCOPEG077.46$+$01.76, and SCOPEG181.92$+$00.36. The first two, SCOPEG015.01$-$00.67 and SCOPEG035.58$+$00.05, are coincident with ATLASGAL sources, with the masses reported here higher and with smaller radii. We found the former to be prestellar, however, the ATLASGAL source is associated with a tracer of massive star formation. These sources close to the border of this region may move into, or further away from it, with a different temperature or distance assignment. However, as candidate young massive clusters, additional data are needed to confirm their nature.

\subsection{Star formation efficiency and $L/M$}
\label{sec:sfe}

\subsubsection{$L/M$ in individual compact sources}
\label{sec:sfe_clumps}

\begin{figure}
\includegraphics[width=0.49\textwidth]{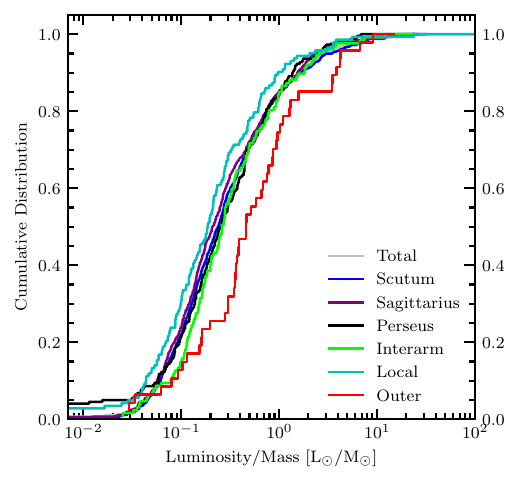}
\caption{Cumulative distributions of the luminosity-to-mass ratio for individual compact sources in the whole PGCC sample and the separate spiral arms. The colours are as described in Fig.~\ref{radii+densities}.}
\label{sfe_histo}
\end{figure}

\begin{table}
\centering
\caption{Statistics for $L/M$ in the Hi-GAL-associated, star-forming compact-source sample (first row) and the individual spiral-arm subsets. The mean ($\bar{x}$), standard error, standard deviation ($\sigma$), median, and Anderson--Darling $p$-values are given.}
\label{sfe_stats}
\begin{tabular}{lccccc}
\hline
Spiral Arm & $\bar{x}$ & $\frac{\sigma}{\sqrt{N}}$ & $\sigma$ & Median & A--D\\
& & & & & $p$-value \\
\hline
{\bf Total} & 0.73 & 0.05 & 2.09 & 0.24 & \\
Scutum--Centaurus & 0.80 & 0.06 & 2.00 & 0.24 & $\gg$ 0.250 \\
Sagittarius & 0.75 & 0.09 & 2.14 & 0.21 & $\gg$ 0.250\\
Perseus & 0.71 & 0.12 & 1.84 & 0.26 & $\gg$ 0.250 \\
Local & 0.62 & 0.18 & 2.14 & 0.19 & 0.029\\
Outer & 1.49 & 0.39 & 2.70 & 0.46 & $\ll$ 0.001 \\
Interarm & 0.90 & 0.22 & 2.50 & 0.27 & 0.209\\
\hline
\end{tabular}
\end{table}

The luminosity-to-mass ratio ($L/M$) is often taken to be an analogue of the instantaneous star-formation efficiency (SFE; e.g., \citealt{Molinari08,Urquhart14,Eden15,Liu2016}), when the luminosity is taken from the embedded young stellar objects and the mass is that of the clump or cloud. This ratio is calculated for the entire star-forming sample  and for the individual spiral-arm subsets of this, using the luminosities from \citet{Elia17,Elia21} and the masses calculated in Section~\ref{sec:masses}. The cumulative distributions of $L/M$ are displayed in Fig.~\ref{sfe_histo}, with the statistics displayed in Table~\ref{sfe_stats}.

The (sub)sample distributions are statistically consistent with each other. However, the mean $L/M$, or SFE, of the Outer spiral arm deviates slightly by $\sim$\,2\,$\upsigma$ from the other arms. The A--D tests demonstrate that the Scutum--Centaurus, Sagittarius and Perseus samples were drawn from the same population, and that the null hypothesis that the Outer arm sources were drawn from the same population can be rejected. The Interarm and Local sources were found to have A--D test $p$-values of 0.209 and 0.029, respectively. The null hypothesis could not be rejected as these values correspond to 2.18\,$\upsigma$ and 1.25\,$\upsigma$, respectively.

These values are significantly lower than those found in other Galactic Plane studies \citep{Eden15,Urquhart18} and are more consistent with the $L/M$ ratio found in quiescent clumps in the ATLASGAL survey, despite the presence of 70-$\upmu$m sources. The values are also consistent with previous SCOPE determinations of $L/M$ \citep{Eden19} of sources located both in and out of the Galactic Plane,  with no evidence found to contradict prior results of low levels of star formation within PGCCs.

\subsubsection{$L/M$ across the Galactic Plane}
\label{SFE_plane}

\begin{figure}
\includegraphics[width=0.49\textwidth]{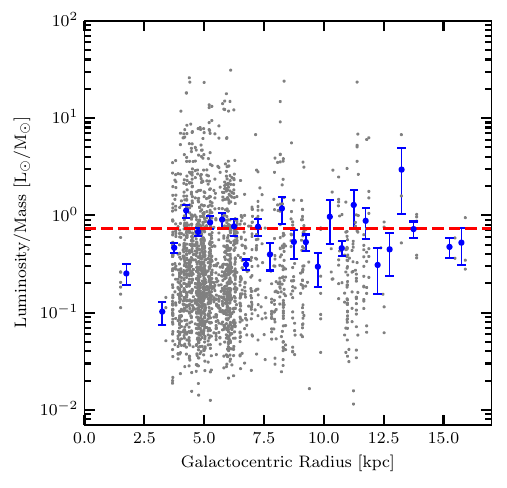}
\caption{$L/M$ as a function of Galactocentric radius. The individual compact sources are displayed by the grey circles. The blue circles are the averages in 0.5-kpc bins, with the associated standard error on the mean. The red horizontal dashed line indicates the Galactic average of 0.73 L$_{\odot}$/M$_{\odot}$.}
\label{sfe_rgc}
\end{figure}

The ratio $L/M$ as a function of Galactocentric radius is shown in Fig.~\ref{sfe_rgc}. This distribution shows the values of individual clumps (grey points) and the mean in each 0.5-kpc bin (blue points). As displayed in Fig.~\ref{sfe_rgc}, there is a large variation in values of $L/M$ from clump to clump. However, the 0.5-kpc averages show very little variation, with only three bins, at 1.75\,kpc, 3.25\,kpc, and 6.75\,kpc, varying from the mean by greater than 5\,$\upsigma$. These bins lie below the average, with those at 1.75\,kpc and 3.25\,kpc containing 7 and 3 compact sources, respectively. The 6.75-kpc bin is consistent with results of \citet{Eden15}, who found a lowered $L/M$ ratio at these Galactocentric radii, coincident with the interarm regions between the Sagittarius and Perseus arms.

The overall trend of no significant variation of the mean on kiloparsec scales is consistent with results from blind Galactic Plane surveys \citep{Moore12,Eden15,Ragan16,Ragan18,Urquhart18}. These studies found no significant increase in $L/M$ that can be attributed to spiral arms, or other features of Galactic structure. Since there is more potentially star-forming material (including PGCCs) in the spiral arms, these studies have concluded that the spiral arms are collecting material, and not impacting the star-formation process.

The similarities to these studies, however, is only when looking at relative changes from bin to bin. The absolute values of $L/M$ are much lower than those reported in similar studies \citep{Eden15,Urquhart18}, as mentioned in Section~\ref{sec:sfe_clumps}. The mean value of $L/M$ in this work is 0.727\,$L_{\odot}/M_{\odot}$, which is almost an order of magnitude lower than those in the aforementioned studies (1.17\,$L_{\odot}/M_{\odot}$ and 18.8\,$L_{\odot}/M_{\odot}$ for \citealt{Eden15} and \citealt{Urquhart18}, respectively). 

\section{Are PGCCs a different population?}

The concentration of star-forming structures in the plane of the Milky Way makes it the ideal laboratory for determining if PGCCs are indeed a separate population.  The previous sections have demonstrated that the physical and star-forming properties of PGCCs (mass, radius, column density, mass surface density, SFE) do not vary as a function of Galactic environment. As a result, we can assume that the whole PGCC catalogue \citep{Planck11,Planck16} can be taken as one population. We can then compare the ATLASGAL dust-continuum-traced clumps that are associated with a catalogued PGCC with those that are not associated with a PGCC, and determine if PGCCs are indeed a separate population.

Positional matching of the PGCC catalogue \citep{Planck11,Planck16} with the ATLASGAL catalogue \citep{Urquhart18} resulted in 2,128 and 5,473 associated and non-associated ATLASGAL-PGCC sources, respectively.

\begin{figure}
\begin{tabular}{l}
\includegraphics[width=0.3\textheight]{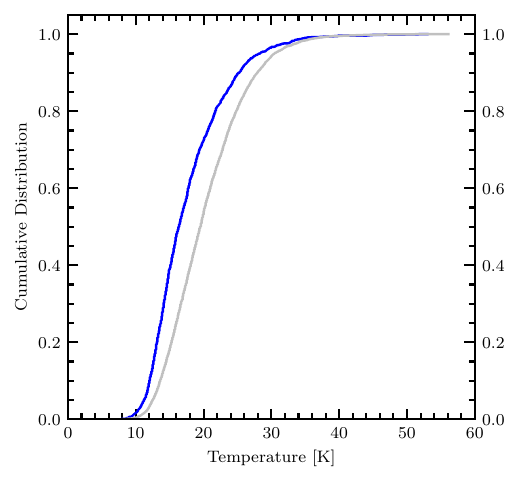} \\ \includegraphics[width=0.3\textheight]{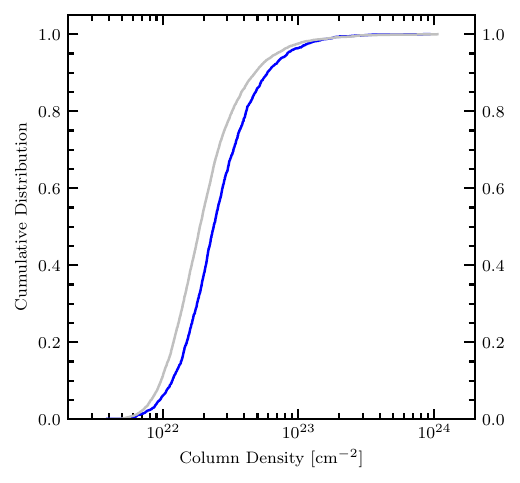} \\
\includegraphics[width=0.3\textheight]{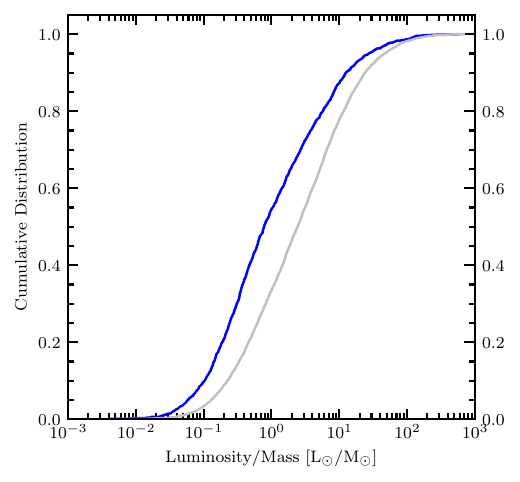} \\
\end{tabular}
\caption{Cumulative distributions of ATLASGAL sources associated with PGCCs (blue) and those not associated (grey). The plots are temperature, column density, and $L/M$ in the top, middle, and bottom panels, respectively.}
\label{atlasgal_physical}
\end{figure}

The ATLASGAL-derived properties of temperature, column density and $L/M$ of these two samples are compared as cumulative distributions in Fig.~\ref{atlasgal_physical}. These properties are chosen due to the cold, dense, low-star-forming nature of PGCCs \citep[e.g.,][]{Planck16,Zhang16,Zhang18}. The distributions in Fig.~\ref{atlasgal_physical} demonstrate that PGCC-ATLASGAL sources are indeed lower in star-formation efficiency, lower in temperature and denser, with A--D tests indicating that they are drawn from different populations (all $p$-values $\ll 0.001$ that they are drawn from the same sample). Further to this, the previous section (Section~\ref{SFE_plane}) determined that the $L/M$ ratio of PGCC-associated compact sources was lower than other Galactic Plane sources, consistent with ratios found for quiescent clumps \citep{Urquhart18}

\section{Summary and Conclusions}

We have investigated the physical properties of $\emph{Planck}$ Galactic Cold Clumps (PGCCs) within one degree of the Galactic Plane from the JCMT Plane Survey (JPS) and SCUBA-2 Continuum Observations of Pre-protostellar Evolution (SCOPE) survey to determine whether PGCCs are a separate population of Galactic clumps.

Using a suite of molecular surveys, velocities, and distances were assigned to 317 PGCCs and 3178 compact sources, also allowing them to be attributed to a spiral arm. These distances were used to determine masses, radii, and mass surface densities for the compact sources, along with column densities.

The mass functions of both the total sample and individual spiral arms are, in general, better described as log-normal distributions as opposed to power laws. All spiral-arm distributions are consistent with each other, except for the Sagittarius Arm and interarm sources, which may reflect the molecular-cloud properties in and out of spiral arms.

After positionally matching the compact sources with young stellar objects from the Hi-GAL survey \citep{Elia17,Elia21}, the sample was split into prestellar and protostellar sources, and the star formation properties were investigated. The radii and column densities of prestellar sources were found to be systematically lower than protostellar clumps. The mass-radius relationships for these two samples agree with each other, and both slopes are consistent with the slopes found in quiescent molecular clouds in an external galactic system \citep{Querejeta21}.

$L/M$ ratios of both individual clumps and globally across the Galactic Plane do not show significant variations as a function of Galactic environment. The mean values found are an order of magnitude lower than those in other Galactic-scale studies, hinting at the lower star-forming content of PGCCs.

Finally, a comparison of ATLASGAL sources \citep{Schuller09,Urquhart18} associated with a PGCC and those not associated was made. The temperatures, column densities, and $L/M$ ratios of the two samples were found to be not drawn from the same population, with lower temperatures, lower $L/M$ ratios, and higher column densities for PGCC-associated sources, confirming PGCCs form a distinct population of Galactic sources.

\section*{Acknowledgements}

This work has been supported by the National Key R$\&$D Program of China (No. 2022YFA1603101). Tie Liu acknowledges the supports by National Natural Science Foundation of China (NSFC) through grants No.12073061 and No.12122307, the international partnership program of Chinese Academy of Sciences through grant No.114231KYSB20200009, and Shanghai Pujiang Program 20PJ1415500. K.T. was supported by JSPS KAKENHI (Grant Number 20H05645). LB and GG gratefully acknowledge support by the ANID BASAL project FB210003. This research was carried out in part at the Jet Propulsion Laboratory, which is operated by the California Institute of Technology under a contract with the National Aeronautics and Space Administration (80NM0018D0004). J.D.F and D.J. are supported by NRC Canada and by individual NSERC Discovery Grants. M.J. acknowledges support from the the Academy of Finland grant No. 348342. K.P. is a Royal Society University Research Fellow, supported by grant no. URF$\backslash$R1$\backslash$211322. C.W.L. is supported by the Basic Science Research Program through the National Research Foundation of Korea (NRF) funded by the Ministry of Education, Science and Technology (NRF-2019R1A2C1010851), and by the Korea Astronomy and Space Science Institute grant funded by the Korea government (MSIT) (Project No. 2023-1-84000). The work of M.G.R. is supported by NOIRLab, which is managed by the Association of Universities for Research in Astronomy (AURA) under a cooperative agreement with the National Science Foundation. P.S. was partially supported by a Grant-in-Aid for Scientific Research (KAKENHI Number JP22H01271 and JP23H01221) of JSPS. G.J.W. gratefully thanks The Leverhulme Trust for an Emeritus Fellowship.

The James Clerk Maxwell Telescope is operated by the East Asian Observatory on behalf of The National Astronomical Observatory of Japan; Academia Sinica Institute of Astronomy and Astrophysics; the Korea Astronomy and Space Science Institute; Center for Astronomical Mega-Science (as well as the National Key R\&D Program of China with No. 2017YFA0402700). Additional funding support is provided by the Science and Technology Facilities Council of the United Kingdom and participating universities in the United Kingdom and Canada. The James Clerk Maxwell Telescope has historically been operated by the Joint Astronomy Centre on behalf of the Science and Technology Facilities Council of the United Kingdom, the National Research Council of Canada and the Netherlands Organisation for Scientific Research. Additional funds for the construction of SCUBA-2 were provided by the Canada Foundation for Innovation. The National Radio Astronomy Observatory is a facility of the National Science Foundation operated under cooperative agreement by Associated Universities, Inc. The authors wish to recognize and acknowledge the very significant cultural role and reverence that the summit of Maunakea has always had within the indigenous Hawaiian community. We are most fortunate to have the opportunity to conduct observations from this mountain This research made use of the data from the Milky Way Imaging Scroll Painting (MWISP) project, which is a multi-line survey in 12CO/13CO/C18O along the northern Galactic Plane with PMO-13.7m telescope. We are grateful to all the members of the MWISP working group, particularly the staff members at PMO-13.7m telescope, for their long-term support. MWISP was sponsored by National Key R\&D Program of China with grant 2017YFA0402701 and CAS Key Research Program of Frontier Sciences with grant QYZDJ-SSW-SLH047. This research has made use of NASA's Astrophysics Data System. The Starlink software \citep{Currie14} is currently supported by the East Asian Observatory.

\section*{Data Availability}

The SCOPE survey and its data products are described fully in \citet{Liu18} and \citet{Eden19}. The JPS and its images and source catalogue can be found in \citet{Eden17}. The associated download instructions are within those papers, or per request to the lead author.

The proposal IDs for these projects can be used to download raw observation data (both surveys) or reduced data products (SCOPE only) from the Canadian Astronomy Data Centre's JCMT Science Archive. The JPS proposal ID is MJLSJ02, whilst the SCOPE survey data can be found with: MJLSY14B, M15AI05, M15BI06, and M16AL003.

\bibliographystyle{mnras}
\bibliography{SCOPE_GP}

\clearpage
\onecolumn

\noindent
Author affiliations:\\
$^{1}$Armagh Observatory and Planetarium, College Hill, Armagh, BT61 9DB, UK\\
$^{2}$Shanghai Astronomical Observatory, Chinese Academy of Sciences, 80 Nandan Road, Shanghai 200030, People's Republic of China\\
$^{3}$Astrophysics Research Institute, Liverpool John Moores University, Liverpool Science Park, iC2, 146 Brownlow Hill. Liverpool, L3 5RF, UK\\
$^{4}$NRC Herzberg Astronomy and Astrophysics, 5071 West Saanich Rd, Victoria, BC V9E 2E7, Canada\\
$^{5}$Department of Physics and Astronomy, University of Victoria, Victoria, BC V8W 2Y2, Canada\\
$^{6}$Jodrell Bank Centre for Astrophysics, School of Physics and Astronomy, The University of Manchester, Oxford Road, Manchester M13 9PL, UK\\
$^{7}$Korea Astronomy and Space Science Institute, 776 Daedeokdae-ro, Yuseong-gu, Daejon 34055, Republic of Korea\\
$^{8}$University of Science and Technology, Korea (UST), 217 Gajeong-ro, Yuseong-gu, Daejeon 34113, Republic of Korea\\
$^{9}$National Astronomical Observatories, Chinese Academy of Sciences, Beijing, 100012, China\\
$^{10}$Key Laboratory of Radio Astronomy, Chinese Academy of Science, Nanjing 210008, China\\
$^{11}$Academia Sinica Institute of Astronomy and Astrophysics, 11F of AS/NTU Astronomy-Mathematics Building, No.1, Sec. 4, Roosevelt Rd, Taipei 10617, Taiwan\\
$^{12}$Department of Physics and Astronomy, University of Calgary, 2500 University Drive NW, Calgary, Alberta T2N 1N4, Canada\\
$^{13}$Nobeyama Radio Observatory, National Astronomical Observatory of Japan,
National Institutes of Natural Sciences,
Nobeyama, Minamimaki, Minamisaku, Nagano 384-1305, Japan\\
$^{14}$Astronomical Science Program,
Graduate Institute for Advanced Studies, SOKENDAI,
2-21-1 Osawa, Mitaka, Tokyo 181-8588, Japan\\
$^{15}$School of Physics and Astronomy, University of Leeds, Leeds, LS2 9JT, UK\\
$^{16}$Department of Astronomy, Peking University, 100871, Beijing China\\
$^{17}$Astronomy Department, Universidad de Chile, Casilla 36-D, Santiago, Chile\\
$^{18}$Department of Physics and Astronomy, James Madison University, MSC 4502, 901 Carrier Drive, Harrisonburg, VA 22807\\
$^{19}$RAL Space, STFC Rutherford Appleton Laboratory, Chilton, Didcot, Oxfordshire OX11 0QX, UK\\
$^{20}$East Asian Observatory, 660 North A'oh\={o}k\={u} Place, Hilo, Hawaii 96720, USA\\
$^{21}$Jet Propulsion Laboratory, California Institute of Technology, 4800 Oak Gove Drive, Pasadena CA 91109, USA\\
$^{22}$Department of Physics, P.O.Box 64, FI-00014, University of Helsinki, Finland\\
$^{23}$Institute of Astronomy, National Tsing Hua University, No. 101, Section 2, Kuang-Fu Road, Hsinchu 30013, Taiwan\\
$^{24}$Center for Informatics and Computation in Astronomy, National Tsing Hua University, No. 101, Section 2, Kuang-Fu Road, Hsinchu 30013, Taiwan\\
$^{25}$Department of Physics, National Tsing Hua University, No. 101, Section 2, Kuang-Fu Road, Hsinchu 30013, Taiwan\\
$^{26}$Department of Physics and Astronomy, University College London, WC1E 6BT London, UK\\
$^{27}$INAF - Osservatorio Astronomico di Trieste, Via G.B. Tiepolo 11, I-34143, Trieste, Italy\\
$^{28}$Gemini Observatory/NSF's NOIRLab, 670 N. A'ohoku Place, Hilo, Hawai'i, 96720, USA\\
$^{29}$National Astronomical Observatory of Japan, National Institutes of Natural Sciences, 2-21-1 Osawa, Mitaka, Tokyo 181-8588, Japan\\
$^{30}$Astronomical Science Program, Graduate Institute for Advanced Studies, SOKENDAI, 2-21-1 Osawa, Mitaka, Tokyo 181-8588, Japan\\
$^{31}$IAPS-INAF, Via Fosso del Cavaliere, 100, I-00133 Rome, Italy\\
$^{32}$Centre for Astrophysics and Planetary Science, University of Kent, Canterbury, CT2 7NH, UK\\
$^{33}$Oststrasse 15, 51674 Wiehl, Germany\\
$^{34}$School of Physical Sciences, The Open University, Walton Hall, Milton Keynes, MK7 6AA, UK\\
$^{35}$National Astronomical Research Institute of Thailand (Public Organization), 260 Moo 4, T. Donkaew, A. Maerim, Chiangmai, 50180, Thailand

\bsp
\label{lastpage}

\end{document}